\documentclass[hidelinks,aps,showpacs,amssymb,superscriptaddress,twocolumn,floatfix]{revtex4-1}
\usepackage{graphicx}
\usepackage{amsmath}
\usepackage{hyperref}
\usepackage{dcolumn}
\usepackage{bm}
\usepackage{color}
\usepackage[utf8]{inputenc}
\usepackage[capitalize]{cleveref}
\usepackage{enumitem}
\usepackage{mathtools}
\allowdisplaybreaks[1]
\begin{document}


\title{Fast, robust and amplified transfer of topological edge modes on time-varying mechanical chain}

\author{I. Brouzos} 
\affiliation{LAUM, UMR-CNRS 6613, Le Mans Universit\'e, Av. O. Messiaen, 72085, Le Mans, France}

\author{I. Kiorpelidis}
\affiliation{LAUM, UMR-CNRS 6613, Le Mans Universit\'e, Av. O. Messiaen, 72085, Le Mans, France}

\author{F. K. Diakonos}
\affiliation{Department of Physics, University of Athens, 15771, Athens, Greece}

\author{G. Theocharis}
\affiliation{LAUM, UMR-CNRS 6613, Le Mans Universit\'e, Av. O. Messiaen, 72085, Le Mans, France}


\begin{abstract}
We show that it is possible to successfully, rapidly and robustly transfer a topological vibrational edge mode across a time-varying mechanical chain. The stiffness values of the springs of the chain are arranged in an alternating staggered way, such that we obtain a mechanical analog of the quantum Su-Schrieffer-Heeger model which exhibits a non trivial topological phase. Using optimal control methods, we are able to design control schemes for driving the stiffness parameters, such that the transfer is done  with high fidelity, speed and robustness against disorder as well as energy amplification of the target edge mode. 
\end{abstract}


\maketitle


\section{Introduction}

The discovery of topological insulators~\cite{Qi2011} in condensed matter physics expands rapidly into many other fields of physics. The example of mechanical metamaterials~\cite{Bertoldi2017, Christensen2015, Surjadi2019} shows the impact of this discovery with the realization of classical mechanical analogs of topological systems~\cite{Susstrunk2015, Nash2015,Ma2019}. Moreover, designing the topology of advanced materials as a tool to control the energy flow or other properties of the mechanical structure, has recently attracted a lot of interest~\cite{Rosa2019, Grinberg2020, Riva2020}.

In the meanwhile, recent studies have explored strategies of exploiting topological properties for the state transfer of localized excited states in quantum systems~\cite{Asboth2016, Boross2019, Mei2018, Longhi2019a, Longhi2019b, Estarellas2017, Lang2017}, a process of great importance for quantum technologies. The inspiration of most works connecting state-transfer and topology is the concept of  Thouless adiabatic quantized pumping~\cite{Thouless1983},  which is based on an adiabatic cyclic modulation of the one-dimensional (1D) potential parameters. It has been realized in several platforms such as semiconductor quantum dots~\cite{Switkes1999, Blumenthal2007}, cold atoms~\cite{Lu2016, Nakajima2016, Lohse2016, Lohse2018}, photons~\cite{Kraus2012, Verbin2015, Zilberberg2018}, artificial spin systems~\cite{Schroer2014, Ma2018} and mechanical materials~\cite{Rosa2019, Grinberg2020}. The key advantage  topology offers in such processes is the inherent protection of boundary edge modes lying in the band gap of dispersion relation when the bulk is topologically non-trivial. This results to a degree of  robustness against disorder. However, in most of these works, topological protection is accompanied with adiabaticity requirements and therefore demands slow processes with sufficiently long total times of evolution in order to avoid non-adiabatic excitations to the bulk states. These are either too slow for the intended application or subject to decoherence effects in quantum and several types of energy loss in mechanical materials.

The scope of this work is to show how to control the energy flow in order to substantially speed up the transfer, with high fidelity and robustness against disorder. A bonus feature of our optimal control schemes is the possibility of amplification of the final target mode, in the sense that more energy is accumulated in this mode by the end of the process compared to its initial energy. We obtain these targets by exploiting the robustness of topologically protected edge modes, the toolbox of optimal control and the possibility to accumulate energy on specific vibrational modes. Let us underline that the latter is prohibited in quantum systems due to unitarity, while it is an extremely useful property in acoustic metamaterials for instance, in order to compensate for inherent losses.  


\section{Time-varying mechanical dimer chain}

\begin{figure}[h!]
\begin{center}
\includegraphics[width=1.0\columnwidth]{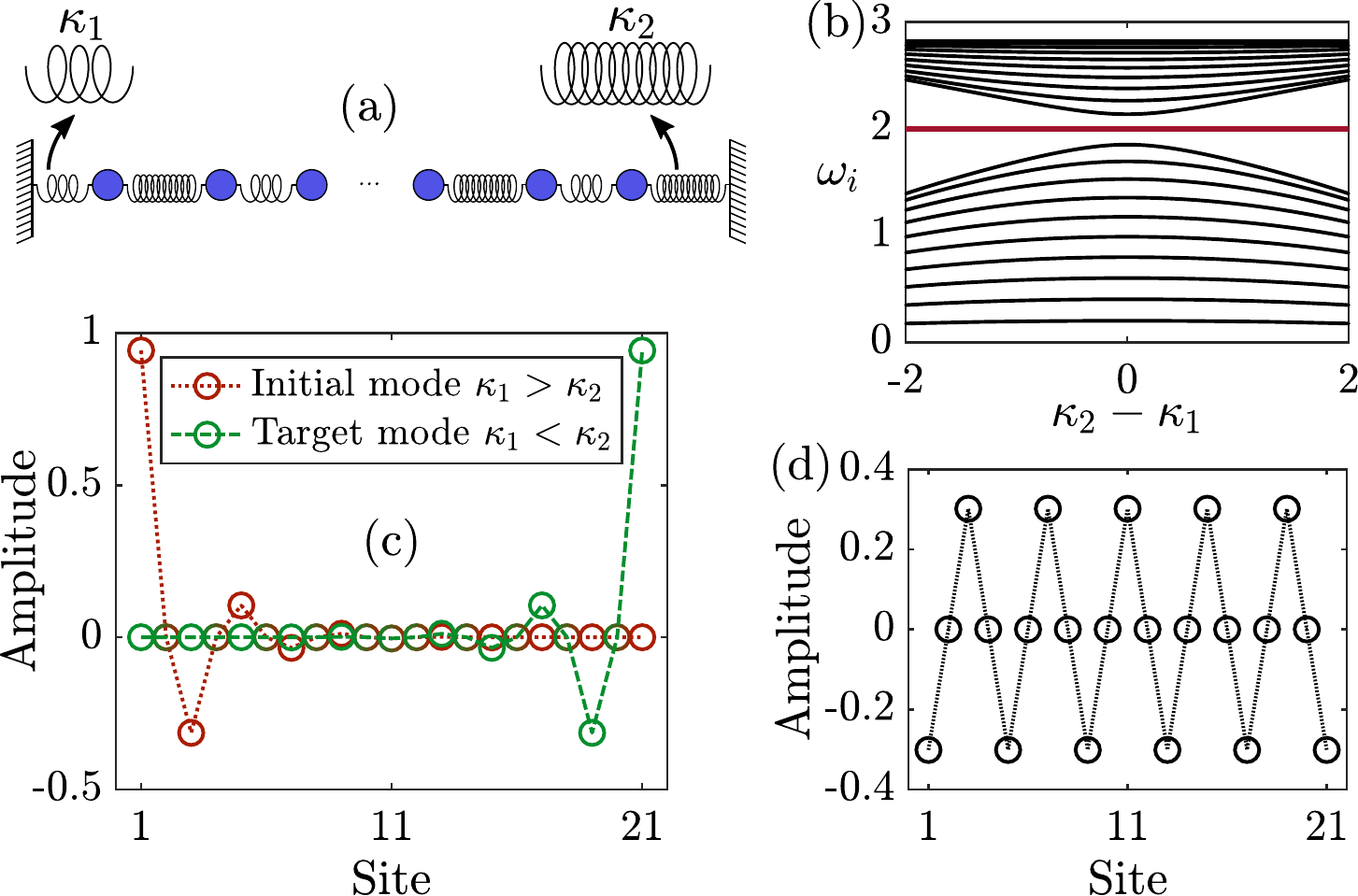}
\end{center}
\caption{(a) Sketch of classical SSH chain with an odd number of masses connected with springs with alternating stiffness $\kappa_1$, $\kappa_2$ and fixed ends. (b) Eigenfrequencies as a function of $\kappa_2-\kappa_1$ keeping $\overline{\kappa}=4$. The edge mode's frequency is kept constant. (c) Amplitudes of initial and target edge mode supported by the system when $\kappa_1>\kappa_2$ and $\kappa_2>\kappa_1$ respectively. We show how with optimal time modulation of $\kappa_1(t)$ and $\kappa_2(t)$ we can transfer from initial to the target mode with high fidelity in very short time. (d) Profile of the edge mode at the closed gap point.}
\label{fig1}
\end{figure}

Our model is the benchmark classical system of N-coupled, single degree of freedom oscillators, with alternating values of the spring stiffnesses $\kappa_1$ and $\kappa_2$, see Fig.~\ref{fig1}(a). It is a relevant model for phonons dynamics and various types of mechanical crystals and metamaterials, as for instance granular systems \cite{Chaunsali2017}, phononic lattices~\cite{Rosa2019} and magneto-mechanical structures~\cite{Grinberg2020} among others. The Lagrangian of small vibrations of this dimer chain is given by
\begin{equation}
\label{lagrangian}
L=\frac{1}{2} \boldsymbol{\dot{q}}^T \mathbf{M} \boldsymbol{\dot{q}} - \frac{1}{2} \boldsymbol{q}^T \mathbf{K} \boldsymbol{q},
\end{equation}
where $\boldsymbol{q}$ is the vector of displacements of size $N$ equal to the number of masses and $\mathbf{M}=m\mathbf{I}_N$ is the mass matrix. $\mathbf{K}$ is the stiffness matrix, given by
\begin{equation}
\label{kmatrix}
\mathbf{K}=~
\begin{bmatrix}
\overline{\kappa}   & -\kappa_2 &0 &  & \dots  \\
-\kappa_2     & \overline{\kappa} &  -\kappa_1 & 0 & \dots \\
0 & -\kappa_1      & \overline{\kappa} &  -\kappa_2 & \dots \\
& &   \dots & & \\
& & \dots     & -\kappa_2 & \overline{\kappa} &  -\kappa_1 \\
&     & \dots & 0 & -\kappa_1 & \overline{\kappa}
\end{bmatrix}
\end{equation}
where $\overline{\kappa} =\kappa_1+\kappa_2$. Note, as it is depicted in Fig.~\ref{fig1}(a), that fixed boundary conditions are used. The alternating value of the spring stiffness  $\kappa_1$ and $\kappa_2$ induces topological features and, as we will see below, guarantees the existence of robust edge modes.

Research on time-dependent elastic structures has received considerable attention over the last three years. Theoretical studies revealed that time modulations of the material's constitutive properties could be induced in various materials including photosensitive~\cite{Gump2004}, magneto-elastic~\cite{Ansari2017}, piezoelectric~\cite{Croenne2017}, or phase changed chalcogenide~\cite{Lencer2008}. In addition, various experimental platforms with temporal modulated elastic structures have been developed the last two years~\cite{Croenne2017,Trainiti2019,Marconi2020}. These developments allow us  to consider in this work, the spring stiffnesses as time-dependent functions: $\kappa_1=\kappa_1(t)$ and  $\kappa_2=\kappa_2(t)$.

\subsection{Eigenspectrum and edge modes}

The Lagrangian (\ref{lagrangian}) leads to the equations of motion
\begin{equation}
m \boldsymbol{\ddot{q}}=-\mathbf{K}(t) \boldsymbol{q} .
\label{spectrum1}
\end{equation}
For the case of time-independent spring stiffnesses, looking for solutions 
$\boldsymbol{q}(t) \propto e^{i\omega_i t} \boldsymbol{q_i}$, we obtain
\begin{equation}
 m \omega_i^2 \boldsymbol{q_i}= \mathbf{K} \boldsymbol{q_i} ,
\label{spectrum2}
\end{equation}
where $\omega_i$ are the eigenfrequencies and $\boldsymbol{q_i}$ the eigenmodes of the mechanical model.  The index $i=1,...,N$ runs in ascending order of discrete eigenfrequencies, while we use an odd-sized chain with $N=21$ identical masses throughout this work. In Fig.~\ref{fig1}(b), we show the eigenfrequencies $\omega_i$ of $\mathbf{K}$ as a function of the difference  $\kappa_2-\kappa_1$ keeping a constant $\overline{\kappa}=4$ in arbitrary units $\kappa_0$. 

The mechanical dimer chain has chiral symmetry~\cite{Susstrunk2016} which results, for the odd-sized finite chains, to the existence of one edge mode localized at one side of the chain with frequency equal to $\tilde{\omega}=\sqrt{\overline{\kappa}/m}$. It is well separated from the bulk modes which, due to finite size of the system, do not form continuous bands. Moreover, the edge mode has a left (right) localized profile of amplitudes of the corresponding eigenvector $\boldsymbol{\tilde{q}}$ if $\kappa_1>\kappa_2$ ($\kappa_2>\kappa_1$) as shown in Fig.~\ref{fig1}(c) with red dotted (green dashed) line. Hereafter, every variable or parameter with a tilde corresponds to the edge mode. Length unit is the normalization length of the edge mode amplitude vector $L=|\boldsymbol{\tilde{q}} | $ and therefore energy unit is $\kappa_0 L^2$ and time unit $\sqrt{m/\kappa_0}$. In the following, without loss of generality we set $\kappa_0=1$. 

This profile of the edge modes holds as soon as $\kappa_1 \ne \kappa_2$ and is more localized the more these two values of stiffness differ. As we approach $\kappa_1=\kappa_2$, a point in the stiffness parameter space which we will call {\it closed gap point}, the frequencies of the bulk modes approach the edge mode's frequency, as shown in Fig.~\ref{fig1}(b). Exactly at this point the profile of the edge mode, shown in Fig.~\ref{fig1}(d), does not differ qualitatively from that of the bulk modes. However, due to the finite size of the system, the difference between the frequencies of bulk and edge modes is not vanishing at the closed gap point and it decays algebraically to zero for $N \rightarrow \infty$.


\subsection{Energy}

Introducing the conjugate momenta $p_i \equiv \partial L / \partial \dot{q}_i$ with $i=1,...,N$ we obtain the Hamiltonian
\begin{equation}
\label{hamiltonian1}
\mathcal{H}(\boldsymbol{q},\boldsymbol{p},t)=\frac{1}{2}(\boldsymbol{p}^T\boldsymbol{p}+\boldsymbol{q}^T\mathbf{K}(t)\boldsymbol{q}) ,
\end{equation}
describing the phase-space evolution of the considered system. Since $\mathbf{K}(t)$ is a symmetric matrix of positive type, it can be diagonalized  by an orthogonal matrix, the {\it instantaneous modal matrix} $\mathbf{A}(t)$, which is composed by the {\it instantaneous eigenvectors} $\boldsymbol{q_i}(t)$ of $\mathbf{K}(t)$. 
It also holds that $\mathbf{A}(t)^T\mathbf{K}(t)\mathbf{A}(t)=\mathbf{\Delta}(t)$ where $\mathbf{\Delta}(t)=\mathbf{diag}(\omega_1^2(t),...,\omega_N^2(t))$ is the diagonal matrix with elements the {\it instantaneous eigenfrequencies}. With the use of the modal matrix, we can now change the variables $(\boldsymbol{q},\boldsymbol{p})$ to normal variables $(\boldsymbol{Q},\boldsymbol{P})$, given by
\begin{equation}
\boldsymbol{Q}=\mathbf{A}^T\boldsymbol{q},   \boldsymbol{P}=\mathbf{A}^T\boldsymbol{p}.
\label{transf}
\end{equation}
Using the above transformation we can rewrite the Hamiltonian in Eq.~(\ref{hamiltonian1}), keeping in mind that the new variables $\boldsymbol{Q}$, $\boldsymbol{P}$ are not phase space variables for $\mathcal{H}$ any more, since they do not obey the appropriate Poisson bracket conditions with $\mathcal{H}$.
\begin{equation}
\label{hamiltonian2}
\mathcal{H}=\frac{1}{2}(\boldsymbol{P}^T\boldsymbol{P}+\boldsymbol{Q}^T\mathbf{\Delta}(t)\boldsymbol{Q}),
\end{equation}
and Eq.~(\ref{hamiltonian2}) can be then used for writing the total energy as a sum of the {\it modal energy contribution} $E_i$,
\begin{equation}
\label{modalenergy}
E=\sum_{i=1}^N{E_i}=\sum_{i=1}^N\left(\frac{{{P_i}}^2}{2m}+\frac{1}{2}{m\omega_i}^2Q_i^2\right).
\end{equation}
Notice that the values of $E_{tot}$ and $E_{i}$ vary with time and therefore the initial energy $E(0)$ may not be the same as the final one $E(T)$, resulting in an energy loss or gain. Of course the same holds for each modal energy contribution $E_i$ allowing us to refer to amplification of the target edge mode in case that $\tilde{E}(T)>\tilde{E}(0)$.

At this point it is worth noting that the variable change given in Eq.~(\ref{transf}) is canonical, while the generating function of this transformation
\begin{equation}
\label{generating}
F(\boldsymbol{p},\boldsymbol{Q},t)=-\boldsymbol{p}^T \mathbf{A} \boldsymbol{Q},
\end{equation}
suggests that the Hamiltonian $\mathcal{H}'(\boldsymbol{Q},\boldsymbol{P},t)$ in the normal modes representation, i.e., the Hamiltonian that describes the system's dynamics in terms of the normal variables $\boldsymbol{Q}$, $\boldsymbol{P}$ and time $t$, reads 
\begin{equation}
\mathcal{H}'(\boldsymbol{Q},\boldsymbol{P},t)=\frac{1}{2} \left( \boldsymbol{P}^T\boldsymbol{P}+\boldsymbol{Q}^T\mathbf{\Delta}(t)\boldsymbol{Q} \right) - \boldsymbol{P}^T \mathbf{A}^{T} \mathbf{\dot{A}} \boldsymbol{Q}.
\label{hamiltonian3}
\end{equation}
The first two terms of $\mathcal{H}'$ can be written as a sum of $N$ independent Hamiltonians of harmonic oscillators. However, the matrix $\mathbf{A}^{T} \mathbf{\dot{A}}$ is not a diagonal one and when the initial conditions lead to a single normal mode's oscillation, the last term of $\mathcal{H}'$ is causing excitations to other normal modes.


\subsection{Mechanical vs Quantum topological chain}

The under study classical chain acts as an analog of the Su-Schrieffer-Heeger model (SSH)~\cite{Su1979}. The latter was initially conceptualized as a diatomic linear quantum system, with next-neighbor interaction, exhibiting topological properties \cite{Asboth2016}. The  Hamiltonian of this quantum model is similar to the $\mathbf{K}$ matrix canceling the diagonal terms: $\mathbf{H}_{SSH}=\mathbf{K}-\overline{\kappa}\mathbf{I}$. 

There are nevertheless some important differences between the quantum and classical case.  The quantum system evolves according to the Schr\"odinger equation $i\hbar\frac{d\boldsymbol{\psi}}{dt}=\mathbf{H}_{SSH}\boldsymbol{\psi}$ with $\boldsymbol{\psi}=(\psi_1,...,\psi_N)$, which is first order in time, while the classical system follows Newton's equation $\frac{d^2\boldsymbol{q}}{dt^2}=-\mathbf{K}\boldsymbol{q}$. Moreover, the relevant eigenvalue-eigenvector problem   for the quantum case is that of the Hamiltonian operator: $\mathbf{H}_{SSH}\boldsymbol{\psi_i}=\mathcal{E}_i\boldsymbol{\psi_i}$ where $\boldsymbol{\psi_i}$ and $\mathcal{E}_i$ are the eigenvectors and eigenenergies of the SSH Hamiltonian, whereas, for the classical case, it is that of the $\mathbf{K}$ matrix [Eq.~(\ref{spectrum2})].

The K matrix possesses non-vanishing diagonal elements [see Eq.~(\ref{kmatrix})] while in $\mathbf{H}_{SSH}$ the diagonal elements vanish \cite{note1}. Moreover, $\mathbf{H}_{SSH}$ is a hermitian triagonal matrix, implying that when $N$ is odd, an edge state with zero energy is supported and the energies $\mathcal{E}_i$ of the bulk states are also symmetric around zero. However, in the mechanical dimer chain, the form of the off-diagonal part of the $\mathbf{K}$ matrix is similar to $\mathbf{H}_{SSH}$, while the diagonal elements are all equal to $\overline{\kappa}$. Hence, from Eq.~(\ref{spectrum2}) it  follows that: (i) the eigenfrequencies squared of the bulk modes, $\omega_i^2$, are symmetric around $\tilde{\omega}^2$, i.e., the eigenfrequency squared of the edge mode, (b) $\tilde{\omega}^2$ is equal to $\overline{\kappa}$.

The eigenfrequencies squared are also connected to each \textit{normal mode's eigenenergy} $\mathcal{\epsilon}_i$. As explained before, the total energy of the mechanical chain is written as a sum of the modal energy contribution of each mode. Therefore, $\mathcal{\epsilon}_i$ are equal to $\omega_i^2/2$, which are symmetric around $\tilde{\omega}^2/2$. However, the evolution in time of each mode's contribution $E_i$, is proportional to $\exp(2 i \omega_i t)$ implying that the the energy contributions of the bulk modes do not evolve symmetrically in time, since the eigenfrequencies of the bulk modes are not symmetric around $\tilde{\omega}$ (see Appendix \ref{appendixA} for more details).


\section{State transfer via topological chains}

The SSH model has been used recently  in the context of robust state transfer \cite{Estarellas2017, Longhi2019a, Longhi2019b, Mei2018, Lang2017}. The subject of state transfer has attracted major interest over the years, while the concept of quantum state transfer has been accomplished directly via photons~\cite{Bennett1992}, in linear spin chains~\cite{Bose2003}, with quantum dots~\cite{Loss1998} and in many other platforms. However, it was the need for robustness against disorder that led to the use of a non-trivial topological SSH chain with alternating couplings instead of the trivial homogeneous spin chain or engineered coupling chain \cite{Bose2003}.

There are several protocols based on either odd-sized or even-sized SSH chains. For the latter case discussed in \cite{Estarellas2017, Longhi2019b, Lang2017}  the edge states are two and localized in either side of the chain. None of them is an eigenstate of the system: they are both actually a superposition of symmetric and anti-symmetric eigenstates lying close to zero energy (but not exactly vanishing). Therefore, there is a Rabi coupling among the left and right side localized edge states which is very small but can be used for transfer among them \cite{Estarellas2017}. There are also methods to accelerate this transfer protocol~\cite{Longhi2019b, Lang2017}. Moreover, for the even-sized chain,  we have the adiabatic Thouless pumping scenario, which is performed under the Rice-Mele model. The latter  is a modification of the SSH with diagonal terms that can induce a pumping cycle of Thouless-type both in quantum and mechanical case \cite{Asboth2016, Lu2016, Nakajima2016, Lohse2016,Longhi2019a, Longhi2019b, Grinberg2020}.

In this study, we focus on the odd-sized chain scenario like in \cite{Mei2018, Longhi2019a}. This choice is mainly made because we aim at an edge mode which is an eigenmode of the system initially and finally independently from our choice of $\kappa_1$ and $\kappa_2$ parameters. Additionally, we ensure that our initial and target edge modes do not have leaks to other modes and thus we can accumulate and store energy. Since the initial excitation is an eigenmode [see Fig.~\ref{fig1}(c) red dotted line] a time-change of the parameters is needed to drive the system (Rabi-type processes do not occur here).

The purpose of our study is to transfer the energy from the left-localized mode [Fig.~\ref{fig1}(c) red dotted line] to the right-localized one [Fig.~\ref{fig1}(c) green dashed line]. We try to achieve this by time-dependently modulating the spring stiffness which exchange values between initial and final time, namely 
\begin{equation}
\label{I_F}
\kappa_1(0)=\kappa_2(T)~~~,~~~\kappa_2(0)=\kappa_1(T). 
\end{equation}

The values we chose for initial parameters, $\kappa_1(0)=3$ and $\kappa_2(0)=1$,  result in a well localized edge mode, as shown in Fig.~\ref{fig1}(c). Nevertheless, the localization length of our edge mode is not vanishing, as in most cases of the literature, where the small coupling parameter is considered to be vanishing such that there is only a single site excitation (in our case that would correspond to $\kappa_2(0)=0$). We also have the physical restriction that $\kappa_1(t),\kappa_2(t)>0$ for the whole duration of the process. 

The initial and target edge mode are adiabatically connected, meaning that, if the operation is done in very long time $T \to \infty$, one can in principle assume a path through the instantaneous edge mode $\mathbf{\tilde{q}}(t)$ without any excitation which will drive the system from $\mathbf{\tilde{q}}(0)$ on the left to $\mathbf{\tilde{q}}(T)$ on the right side. When the time is finite there will be typically some (even very small) non adiabatic excitations to instantaneous bulk eigenmodes, i.e., $E_i(t) \ne 0$ for $t>0$ and $i\ne \frac{N+1}{2}$. 

One intensively studied method to overcome non-adiabatic effects and overthrow the adiabatic constraint of a sufficiently long total time needed for a successful process, is shortcut to adiabaticity, also known as counteradiabatic or superadiabatic driving \cite{Odelin2019}. With this method, one finds a control Hamiltonian, added to the initial one, that literally cancels the non-adiabatic excitations. Very recently, the first relevant work on state transfer via a quantum topological chain has appeared \cite{DAngelis2020}, with the control Hamiltonian adding next-to-nearest neighbour interactions.

In this work we refrain ourselves from  adding counter-adiabatic terms to the Hamiltonian. Therefore we subsequently always have (even very small) non-adiabatic excitations. Remaining with the SSH coupling terms, we will show how optimal control schemes \cite{Werschnik2007,Caneva2011} improve the time needed to obtain an almost perfect transfer of energy across the edge modes of the chain. This is done by genuinely controlling the energy flow using efficiently the instantaneous separation of the band gap [see Fig.~\ref{fig1}(b)], a main feature of the topological nature of the system.  Optimal control, which has also been used in state transfer context \cite{Caneva2009}, is based on a proper definition of a quantity usually called fidelity (or cost function-infidelity) that has to be maximized (or minimized) as a function of the control parameters of the system under specific constraints.

We are interested here  on the transfer fidelity of the energy of a vibrational eigenmode to another eigenmode of the final system. At the start of the process this eigenmode may be at any phase of its oscillation and be transferred at any phase too. Moreover, as mentioned before, the complication here is that both initial and target modes are not single site excitations, therefore we can not simply define transfer fidelity as the excitation energy of a single site (mass in our case).  

In order to define a proper fidelity  we need to analyze the energy flow in the mass-spring system. To this end we use the $E_i$'s, in order to define a quantity that is normalized for every time instant:
$C_{i}(t)=\frac{E_i(t)}{E(t)}$. Let us call this quantity {\it modal energy share}. We use the modal energy share of the edge mode (corresponding to index $i=\frac{N+1}{2}$) at final time $T$, namely
\begin{equation}
\label{fidelity1}
F=\tilde{C}(T)=\frac{\tilde{E}(T)}{E(T)} ,
\end{equation}
while $F$ depends in general on the initial phase $\phi_0$, since the initial edge mode is an oscillating eigenmode. Therefore, we consider the initial phase as a free parameter ranging from $[0,2\pi)$ and define \textit{fidelity} as the minimum $F$ in this phase interval (see Appendix \ref{appendixB}),
\begin{equation}
\label{fidelity2}
\mathcal{F}=\min_{\phi_0}{F} .
\end{equation}
 

\section{Results: Speeding up the transfer}

As we explained above, for finite time non-adiabatic excitations will appear in our system and therefore it is not possible to have perfect transfer $\mathcal{F}=1$. We set therefore a lower bound for the fidelity $\mathcal{F}=0.99$ and we search for protocols that minimize the necessary total time to achieve it. Instead of directly showing final numerical results we will go step by step improving the speed of the protocol and highlighting the underlying physics of energy flow management that each protocol does.

We begin our analysis by studying a scheme considered in \cite{Mei2018}, which consists of a single frequency, semi-cyclic trigonometric control function. This scenario belongs to a wider class of protocols, which can be written in the form
\begin{equation}
\label{couplings}
\kappa_1(t) = \kappa^++\kappa^- f(t)~~~,~~~\kappa_2(t) = \kappa^+-\kappa^-f(t),
\end{equation}
where $\kappa^+=\frac{\kappa_1(0)+\kappa_2(0)}{2}$ and $\kappa^-=\frac{\kappa_1(0)-\kappa_2(0)}{2}$, while the specific choice of $f(t)$ for the \textit{trigonometric protocol} is given by
\begin{equation}
\label{cosine}
f(t) = \cos \left( \dfrac{\pi t}{T} \right).
\end{equation}

\begin{figure}[h!]
\includegraphics[width=1\columnwidth]{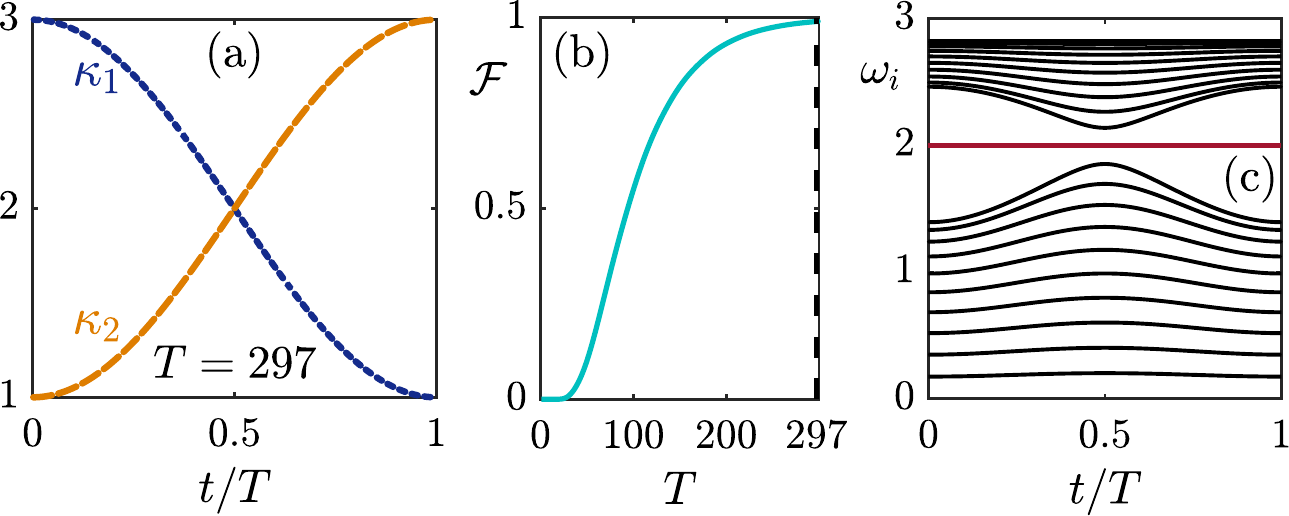}
\caption{(a) $\kappa_1(t)$ and $\kappa_2(t)$ for the trigonometric protocol. (b) $\mathcal{F}(T)$ and (c) $\omega_i(t)$ for the trigonometric protocol.}
\label{fig2}
\end{figure}

This control scheme, shown in Fig.~\ref{fig2}(a), results in a fidelity $\mathcal{F}$ that increases smoothly with $T$ and approaches unity as $T\to \infty$ [see Fig.~\ref{fig2}(b)] while it reaches the target fidelity $\mathcal{F}>99\%$  at $T_\text{trig}=297$. 

The trigonometric protocol has the property that for every $t$, $\overline{\kappa}(t) = \kappa_1(t)+\kappa_2(t)$ is constant and equal to $\kappa_1(0)+ \kappa_2(0)$, resulting in a constant edge mode's frequency during the whole time of the process $\left( \text{recall that~} \tilde{\omega}(t) = \sqrt{\overline{\kappa}(t)} \right)$ as shown in Fig.~\ref{fig2}(c). We will now illustrate why this protocol is not appropriately using the features of the dispersion relation, and what type of control functions improve the total time needed to achieve the target fidelity.

\begin{figure}[h!]
\includegraphics[width=1\columnwidth]{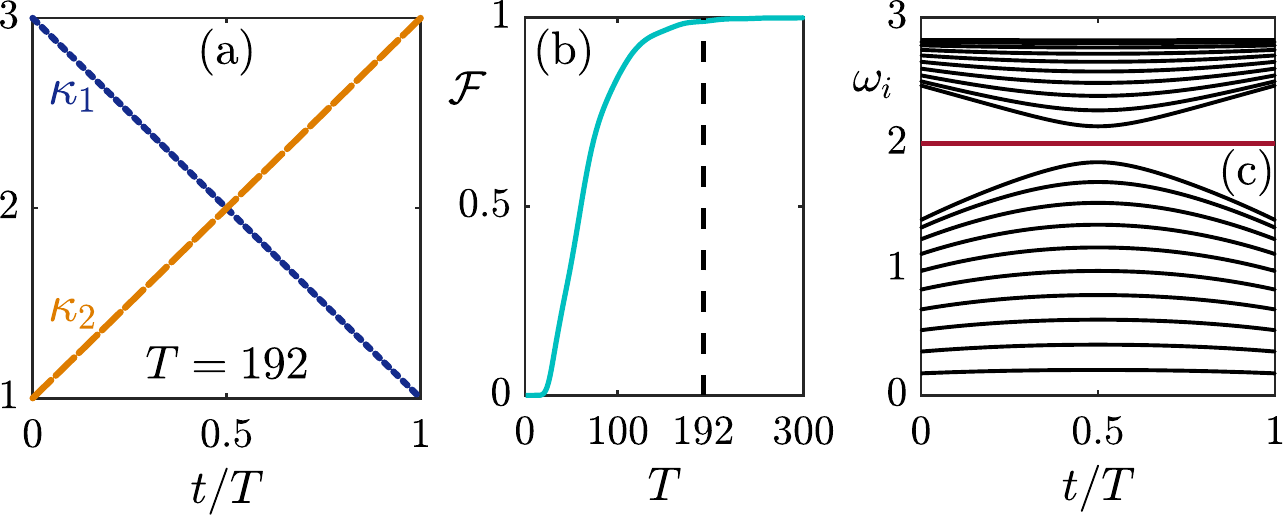}
\caption{(a) $\kappa_1(t)$ and $\kappa_2(t)$ for the linear protocol. (b) $\mathcal{F}(T)$ and (c) $\omega_i(t)$ for the linear protocol.}
\label{fig3}
\end{figure}

To this end, we  consider now a simple protocol, the \textit{linear protocol}, consisting of couplings changing linearly in time [see Fig.~\ref{fig3}(a)], namely
\begin{equation}
\label{linear}
f(t) = 1-2 \dfrac{t}{T}.
\end{equation}
As shown in Fig.~\ref{fig3}(b), this protocol has smoothly increasing fidelity with $T$ and reaches $\mathcal{F}>99\%$ for total time $T_\text{linear}=192$. The improvement we see here in total time needed, compared to the trigonometric protocol, is due to the following fact: The linear function approaches to and departs from the closed gap point at the same rate, while the trigonometric function is slower in the beginning (and in the end) when the gap is very open and fast when it approaches the closed gap point. This can be also seen by comparing Fig.~\ref{fig3}(c) for the linear with Fig.~\ref{fig2}(c) for the trigonometric. The probability of non-adiabatic excitations becomes higher when the gap closes, while it is very suppressed when the topologically induced gap is open. Therefore we conclude that the trigonometric protocol is on the wrong direction of energy flow manipulation, and therefore needs relatively longer times to suppress non-adiabatic excitations and reach the target fidelity than other protocols.

\begin{figure}[h!]
\includegraphics[width=1.0\columnwidth]{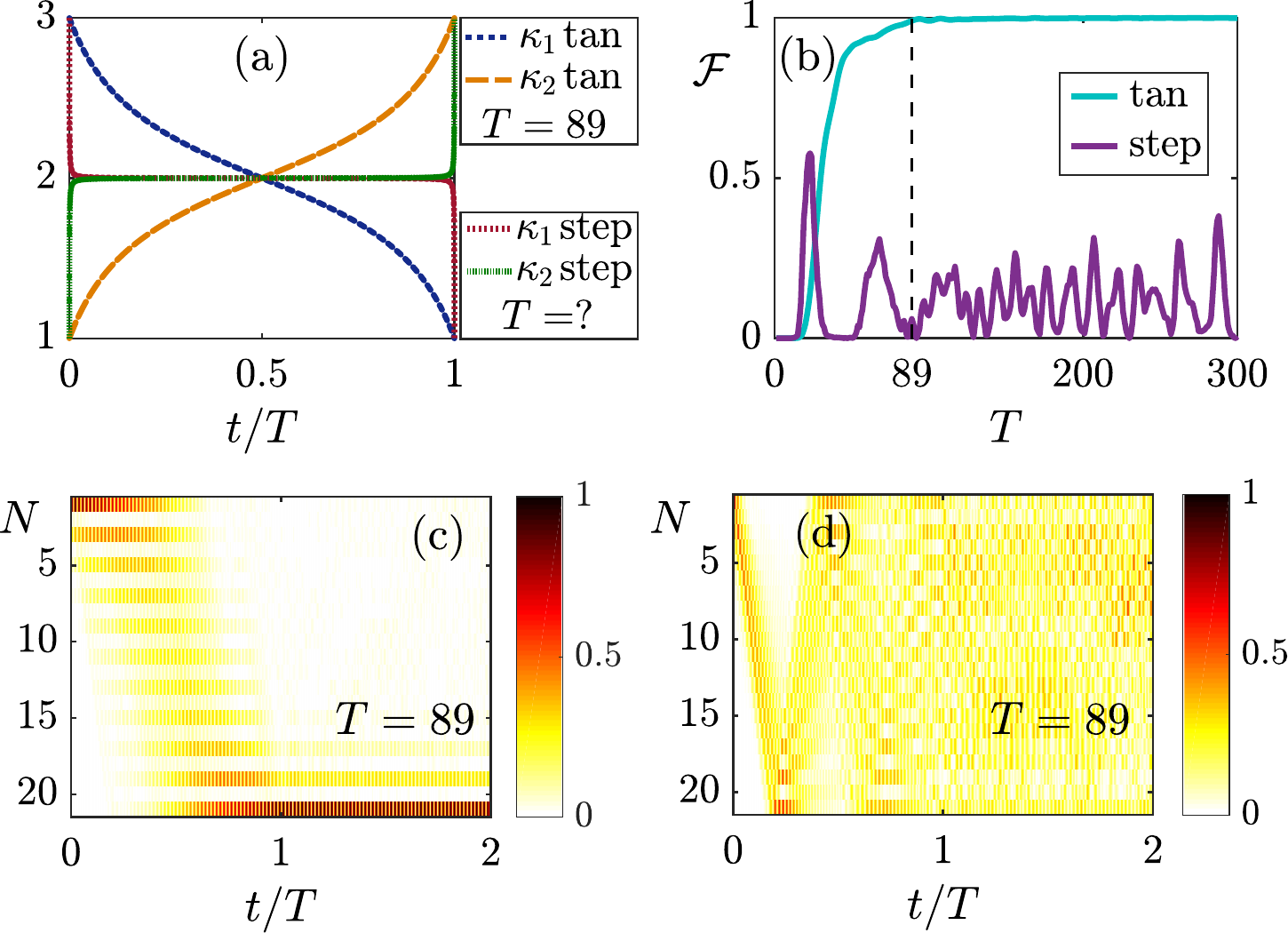}
\caption{(a) $\kappa_1(t)$ and $\kappa_2(t)$ for tangential and step protocols. (b) $\mathcal{F}(T)$ for the corresponding schemes of (a). (c), (d) Time evolution of the absolute value of the particle displacements of the chain for the tangential and step protocols respectively.}
\label{fig4}
\end{figure}

Let us then change the control function in a direction opposite to that of the trigonometric protocol and consider now the scenario of approaching rather fast the closed gap point and staying longer in the vicinity of it. A convenient control function, among others, that can be used to implement and test this scenario is the \textit{tangential protocol}, with a control function given by
\begin{equation}
\label{tangential}
f(t) = \dfrac{\tan(\pi t/T_f +\alpha)}{\tan \alpha}
\end{equation}
where $T_f = \frac{\pi T}{2\pi-2\alpha}$, with free parameter $\alpha \in (\pi/2,\pi)$. With a simple optimization search in the range of the parameter $\alpha$  of the tangential protocol, we find that for the value $\alpha=\pi/2+ 0.4$ [shown in Fig.~\ref{fig4}(a)] the system reaches the desired fidelity at $T_{\text{tan}}=89$ [see Fig.~\ref{fig4}(b)], which is $70\%$ shorter than the trigonometric protocol. Moreover, in Fig.~\ref{fig4}(c) we show  the spatio-temporal evolution of the absolute value of the particle displacements with zero initial phase, meaning zero initial velocities and maximum displacements. With the evolution at a final time $2T$, with $\kappa_{1,2}(T \leq t \leq 2T) = \kappa_{1,2}(T)$, we verify that the edge mode remains localized at the other side of the chain, after reaching the target mode at time $T_{\text{tan}}$.

We stress here that although we have found  a protocol which speeds up the energy transfer by a faster approach of the closed gap point and a longer stay in its vicinity, one cannot assume that this can be taken to the extreme. Indeed, for example, the tangential protocol with $\alpha=\pi/2+0.001$ (close to step function), shown in Fig.~\ref{fig4}(a), does not reach high values of fidelity and $\mathcal{F}(T)$ oscillates strongly as shown in Fig.~\ref{fig4}(b). Furthermore, Fig.~\ref{fig4}(d) illustrating the spatio-temporal evolution at the final time $T=89$, i.e., the final time that the previous scheme reaches the target fidelity, shows that the initial edge mode truly doesn't reach the target one. Let us explain this: The longer the stay at the vicinity of the closed gap the more prominent are the excitation dynamics between the modes, as well as phase oscillations of each mode. Therefore the modal energy contribution at final time $E_i(T)$ depends a lot on the duration of the process with respect to the time scales of these inter- and intra- mode dynamics. In general we conclude that in order to speed up this transfer process one needs to carefully and appropriately use the properties of the dispersion relation of the topological model in a very subtle way.

\begin{figure}[h!]
\includegraphics[width=1\columnwidth]{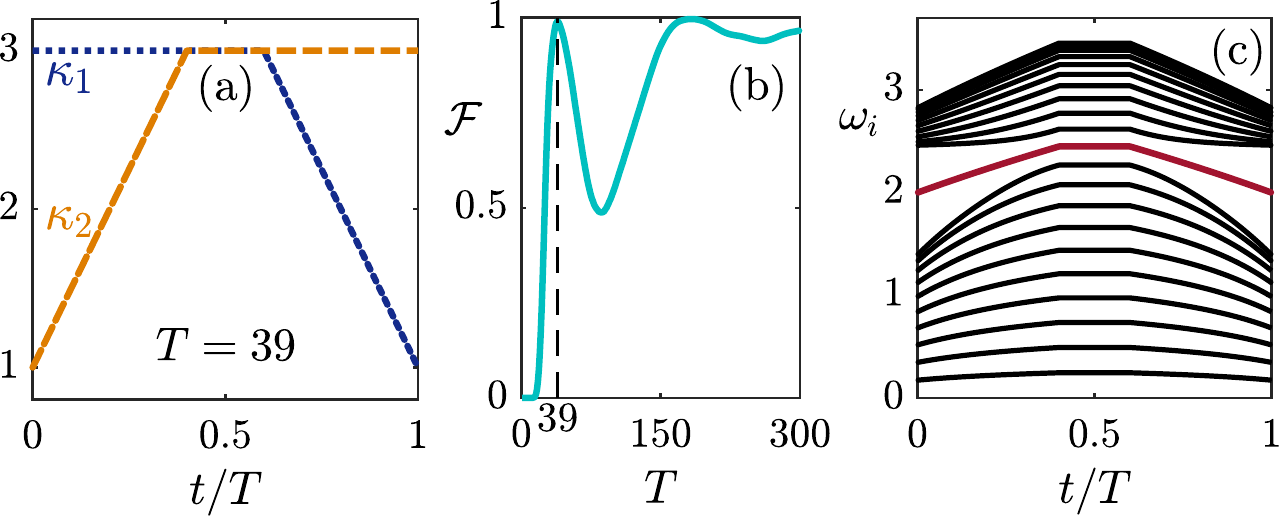}
\caption{(a) $\kappa_1(t)$ and $\kappa_2(t)$ for the 3-step protocol. (b) $\mathcal{F}(T)$ and (c) $\omega_i(t)$ for the 3-step protocol.}
\label{fig5}
\end{figure}

All the aforementioned schemes have the property of keeping the edge mode's frequency constant during the whole process. The question we will try to answer now is whether we could gain some transfer speed if we raise this constraint. We begin this part by studying the \textit{3-step protocol}, a protocol that doesn't keep $\tilde{\omega}$ constant during the process, it is however bounded by the initial value of springs $\kappa_1$ (or the final value of springs $\kappa_2$), meaning $\kappa_{1,2}(t) \leq \kappa_1(0)$ (or $\kappa_{1,2}(t) \leq \kappa_2(T)$) for every $t$ of the process. This control scheme, shown in Fig.~\ref{fig5}(a), consists of 3 time intervals: (i) in the first time interval the large stiffness remains constant to maximum value and the small stiffness increases up to this value, (ii) in the second time interval we have both stiffness constant and equal to their maximum value, (iii) in the third time interval  the large stiffness drops to the small stiffness value and the other one remains constant at the maximum value. In fact, if we keep the mirror symmetry $\kappa_2(t)=\kappa_1(T-t)$ there is only one free parameter: the slope of the linear drive, or equivalently the time interval given for the stiffness to increase (decrease) from initial to final value (and vice versa). Searching the optimal value of this time interval as a portion of the total time we find that $t_{op}=0.4 T$. The corresponding fidelity with total time $\mathcal{F}(T)$, shown in Fig.~\ref{fig5}(b), has oscillating behaviour,  as expected from the time spent near closed gap. Nevertheless the first peak reaches target fidelity $\mathcal{F}>99\%$ at time $T_\text{3-step}=39$ which is 7 times shorter than the  trigonometric scheme. A note on optimization within the above functional constraints is in order here: for both the 3-step and the tangential schemes, even if we raise the mirror symmetric condition $\kappa_2(t)=\kappa_1(T-t)$ and let two free parameters (one for $\kappa_1$ and one for $\kappa_2$), optimization procedure returns values for the corresponding optimized parameters that render back the mirror symmetry. 

But what makes the 3-step protocol faster? It is exactly the idea of not keeping a constant $\overline{\kappa}$, but actually increase the time scales by effectively increasing the mean value of stiffness of the springs during the transfer process. This shifts up and down the value of the instantaneous eigenfrequency $\tilde{\omega}$ as a function of time, as shown in Fig.~\ref{fig5}(c). Therefore, the time scales which may be considered inverse proportional to time average of $\tilde{\omega}$ effectively decrease. 

\begin{figure}[h!]
\includegraphics[width=1\columnwidth]{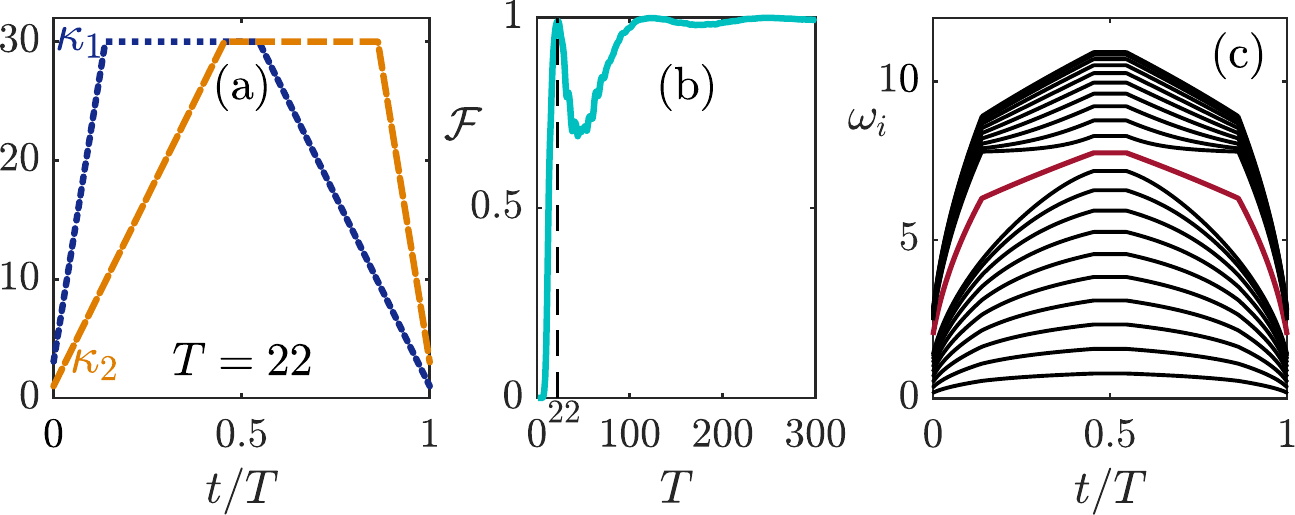}
\caption{(a) $\kappa_1(t)$ and $\kappa_2(t)$ for the 3-step up protocol. (b) $\mathcal{F}(T)$ and (c) $\omega_i(t)$ for the 3-step up protocol.}
\label{fig6}
\end{figure}

Following this line of thoughts, a direct step towards speeding up the transfer further, is to raise up the upper bound of the maximum $\overline{\kappa}$ in the sense that the couplings can exceed the reference value $\kappa_1(0)$ (or $\kappa_2(T)$). Let us take another 3-step protocol, which we refer to as \textit{3-step up protocol}, where in the second time interval the constant maximum value of stiffness is $\kappa_{max}=30$, i.e., one order of magnitude larger than the corresponding value of the previous protocol, as shown in Fig.~\ref{fig6}(a). Both $\kappa_1$ and $\kappa_2$ increase now in the first time interval which may be different for each of them.  Optimization of 4 total parameters now, renders back the mirror symmetry $\kappa_2(t)=\kappa_1(T-t)$ again and therefore constrains the parameter space to two parameters. The time interval in which $\kappa_1$ is constant at value $\kappa_{max}$ is $[0.138T,0.546T]$, suggesting that the corresponding time interval for $\kappa_2$ to be constant at this value is $[0.454T,0.862T]$. Moreover, the fidelity has an oscillating behavior as a function of the total time, as shown in Fig.~\ref{fig6}(b), but it reaches $\mathcal{F}>99\%$ at total time $T_{\text{3-step up}}=22$, which is another significant reduction of $43\%$ compared to the 3-step protocol. 

The qualitative difference of the 3-step up protocol is apparent if one observes the instantaneous eigenfrequency evolution $\omega_i$, as shown in Fig.~\ref{fig6}(c), in comparison to 3-step [Fig.~\ref{fig5}(c)].  The eigenfrequencies change substantially here, while the closed gap occurs at large value of $\kappa_1=\kappa_2=\kappa_{max}=30$. 

\begin{figure}[h!]
\includegraphics[width=1.0\columnwidth]{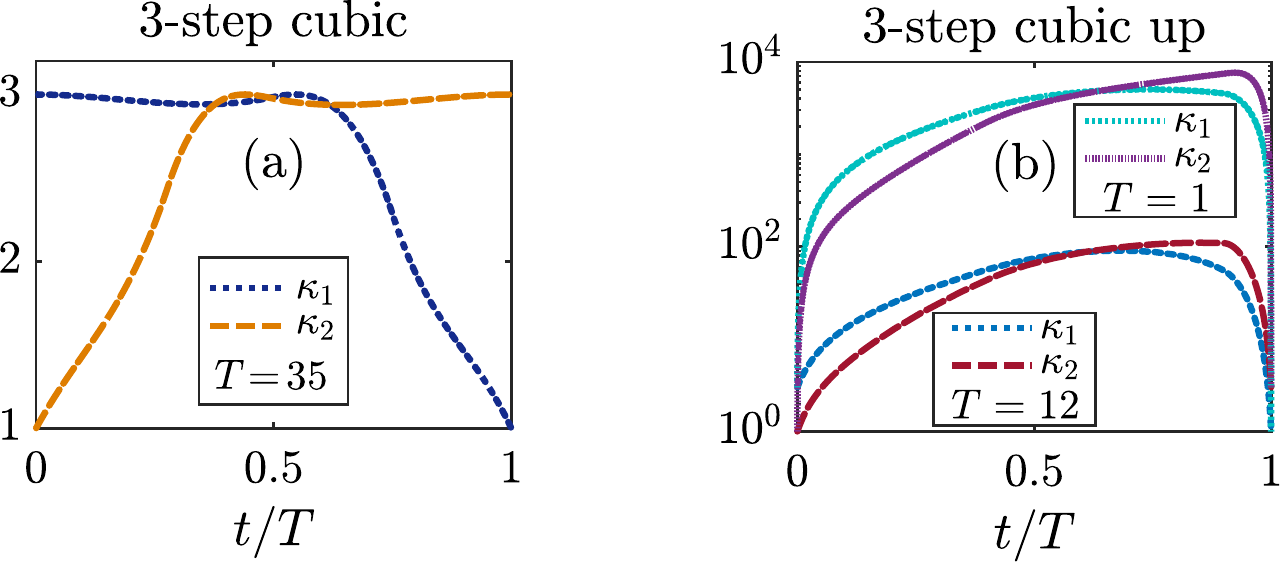}
\caption{(a) $\kappa_{1,2}(t)$ for the 3-step cubic protocol. (b) Corresponding $\kappa_{1,2}(t)$ for the 3-step cubic up protocols.}
\label{fig7}
\end{figure}

Up to now we have gained a lot of speed in the transfer. Further improvement can occur, if we assume other control schemes with functional forms  including more parameters. For instance, we can have  \textit{3-step cubic protocols}/\textit{3-step cubic up protocols} in the sense that each interval of time is a cubic polynomial with adjusted optimal parameters. We have optimized such control schemes involving a parameter space with 16 independent parameters. The protocols found and shown in Fig.~\ref{fig7}  decrease the time even further. Specifically, the 3-step cubic protocol reaches the fidelity $\mathcal{F}>99\%$ at total time $T_\text{3-step cubic}=35$ (recall that $T_\text{3-step}=39$), while for the 3-step cubic up protocol this total time is $T_\text{3-step cubic up}=12$ (recall that $T_\text{3-step up}=22$). We  also report that, if we set the initial phase of the edge mode oscillation to a known value [eg. zero like in Fig.~\ref{fig1}(c)] and not search over all initial phases in order to define fidelity, then we can even find 3-step cubic up protocols that drive to the target mode in total times less than the edge mode period $T_\text{cub}\approx 1<\tilde{T}=\pi$. 

\section{Robustness  and Amplification}

So far, we have seen how to gain a lot of transfer speed with various modifications of the way that the couplings change in time. However, one may be worried whether this improvement in speed is in cost of robustness against disorder due to the presence of fast processes in system's evolution. Next, we study the impact of disorder  on the following representative control schemes: the trigonometric, the tangential, the 3-step,  and the  3-step up  protocols.

We consider a disorder that does not affect the control schemes, but only shifts up and down the initial values of couplings. Its effect is described by
\begin{equation}
\label{dis.stifn}
\kappa_{n}(0) \rightarrow \kappa_n(0) \left( 1 + \epsilon  w_n \right)~,~~~n=1,...,N+1
\end{equation}
with $\epsilon=0.2$ and $w_n \in \left[-1,1\right]$ a uniformly distributed random number. Notice that the system consists of $N+1$ couplings (22 in our case), while when the index $n$ is odd (even), the couplings are of type $\kappa_1$ ($\kappa_2$) [see Fig.~\ref{fig1}(a)]. We also note that the under study disorder, changes the edge mode's frequency, which is no longer equal to $\overline{\kappa}$, since each diagonal element of $\mathbf{K}$ is altered by a different quantity [see Eq.~(\ref{kmatrix})]. Consequently, the system's chiral symmetry is broken. Nevertheless, it is a natural choice, since all couplings are perturbed differently and according to their initial values.

\begin{figure}[h!]
\includegraphics[width=1.0\columnwidth]{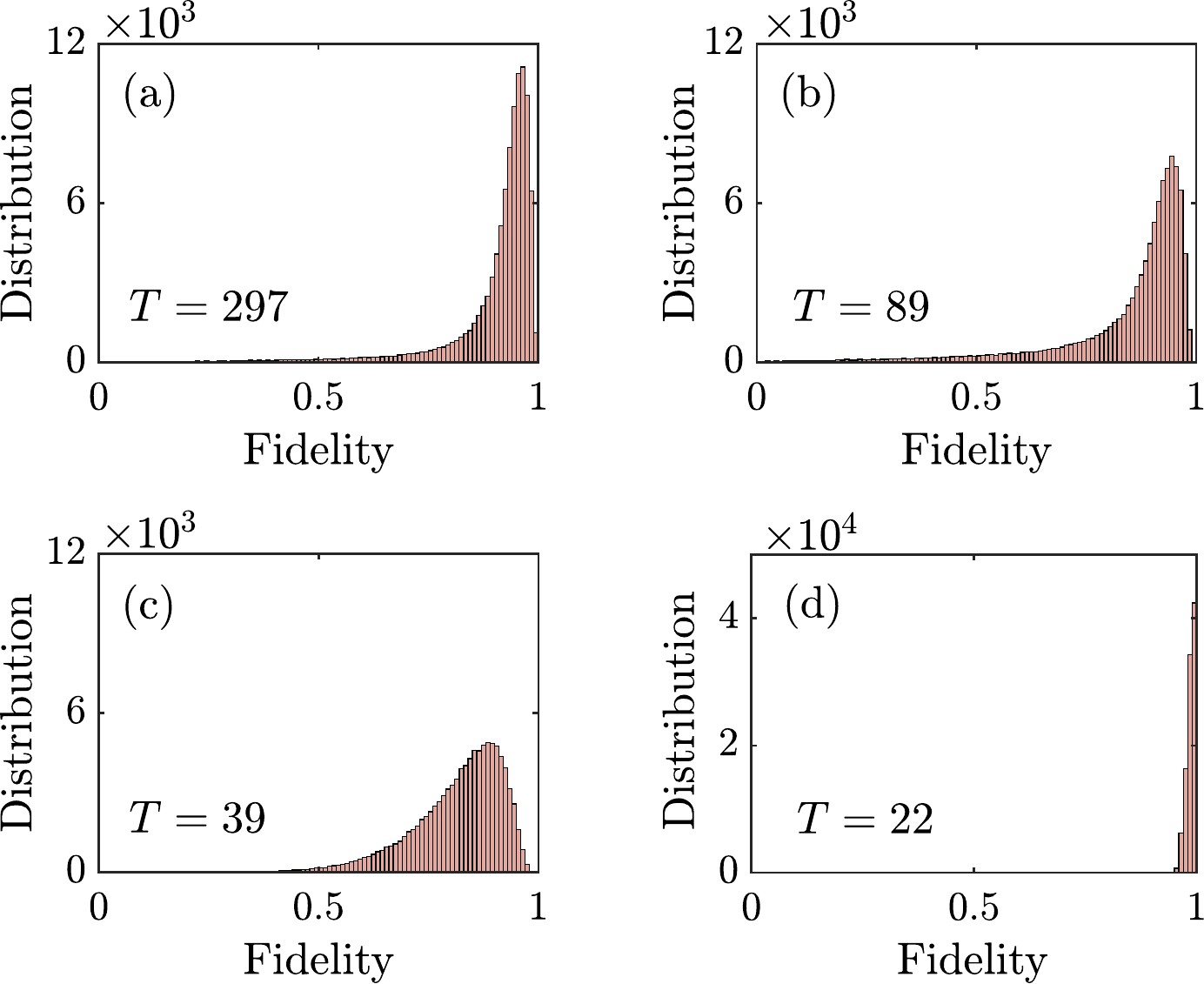}
\caption{Robustness against disorder for 4 protocols and $\epsilon=0.2$. (a) Trigonometric. (b) Tangential. (c) 3-step and (d) 3-step up. Shown is the statistical distribution out of $10^5$ realizations of disorder.}
\label{fig8}
\end{figure}

Without loss of generality we take different realizations of this disorder with vanishing initial phase, calculating $F(\phi_0=0)$ for each realization and show its statistical distribution in Fig.~\ref{fig8}. Note that the total time is fixed for each protocol to the values calculated before, when each realization reaches $\mathcal{F}>0.99$ in order to check weather the improvement in speed comes with a deficit for robustness.

Indeed the trigonometric protocol (with the slowest speed and thus longest time) is apparently outperforming in robustness the tangential and the 3-step. However, the 3-step up is actually even more robust than the trigonometric. This increased robustness is attributed to the fact that the edge mode remains well separated from the bulk throughout the process as we have seen in Fig.~\ref{fig6}(c).

The 3-step up protocol, apart from being the fastest and the most robust, has also an additional bonus property. It can result to amplification, i.e., the final vibrational energy of the target right-localized edge mode can exceed the initial energy of the left-localized edge mode. 

Note that in the classical system there is nothing prohibiting the modal energy contribution of the target edge mode $\tilde{E}(T)$ taking a greater (or lower) value than the energy given at the start of the process, to the initial edge mode [which is also the initial excitation energy given to the system $\tilde{E}(0)=E(0)$]. Therefore, we can define a quantity which we will call {\it energy amplification of the edge mode} as follows
\begin{equation}
\label{harvesting}
A= \frac{\tilde{E}(T)}{\tilde{E}(0)} =  \frac{\tilde{E}(T)}{E(0)} . 
\end{equation}
We additionally note here that the amplification $A$, as well as most of the quantities defined before, depend on the initial phase $\phi_0$. 

\begin{figure}[h!]
\includegraphics[width=0.5\columnwidth]{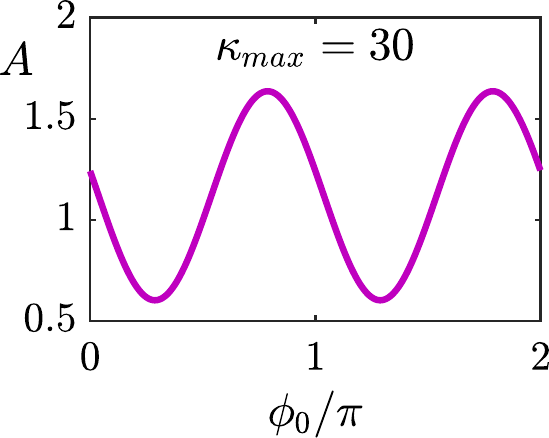}
\caption{Amplification $A$ as a function of the initial phase $\phi_0$ for the 3-step up protocol and when $\kappa_{max}=30$.}
\label{fig9}
\end{figure}

In Fig.~\ref{fig9} we show that we can achieve $A>1$ using the 3-step up protocol [with $k_{max}=30$ as before] for various values of $\phi_0$, but we can also get $A<1$ for several initial phases. The maximum value of $A$ in this case is $1.637$ while  for $\phi_0=0$ we get $A \approx 1.24$. We stress here that the fidelity is above $99\%$ for all of these cases.

Therefore we conclude that  the 3-step up protocol features all desirable properties of state transfer: fastness, robustness and amplification. Of course similar results with even improved such properties can be obtained with protocols like 3-step cubic up that involve more optimized parameters as we have shown. Let us note that all protocols constrained to the initial regime of parameter (trigonometric, linear, tangential, 3-step, 3-step cubic) do not exhibit the feature of amplification.


\section{Conclusions and Perspectives}

We have examined the properties of a mechanical analog of the topological SSH model. We defined properly the quantities to study the energy flow in the system, in order to be able to optimize with respect to the fidelity of the energy transfer on specific vibrational modes. We have designed several protocols that decrease the time needed to transfer the energy from an edge mode of a topological mass spring chain localized on one end to the mirror symmetric one on the other end.  The important qualitative modifications those protocols induce on the dynamics of the system have been explained. We have provided a simple protocol consisted of 3 time intervals with linear increase, constant and linear decrease of the spring stiffness parameter that has both high fidelity, robustness against disorder and energy amplification on the target edge mode. 

The promissing results of our research for robust and rapid control of energy flow in mechanical metamaterials open a lot of questions and further research directions. For instance one may consider the comparison of our optimal control approach with other approaches for speeding up adiabatic processes like transitionless, superadiabatic, shortcuts to adiabaticity. Another perspective is to look deeper at the connection between degree of adiabaticity, the optimal parameters for each functional form of the control schemes (and the symmetries occurring among them) and the energy amplification.  Additionally a lot of the characteristics of the optimal schemes, eg. the mirror symmetry, and the parameter space can unveil further insight about the problem. Last but not least, the  multiple platforms for realization of such transfer processes in experimental setups, may signify additional constraints and possibilities for the control, which could induce further research with respect to the size of the system, the range of the parameters and robustness against other sources of disorder.


\section{Acknowledgements}  

This work has been funded by the project CS.MICRO funded under the program Etoiles Montantes of the Region Pays de la Loire. I. K. acknowledges financial support from Academy of Athens. I. B. and I. K.  contributed equally to this work.


\begin{widetext}

\appendix

\section{Non-symmetric excitation of bulk}
\label{appendixA}

In the quantum SSH system the eigenergies are always symmetric around the zero energy state, i.e., $|\mathcal{E}_i|=|\mathcal{E}_{N-i}|$. In the mechanical system, the stifness matrix $\mathbf{K}$ has a very similar form with $\mathbf{H}_{SSH}$, implying that the eigenvalues of $\mathbf{K}$ are symmetric around the eigenvalue that corresponds to the edge mode. However, the eigenvalues of $\mathbf{K}$ are proportional to the eigenfrequencies squared, therefore $\omega_i^2$ as well as the mode's eigenergies $\epsilon_i$ which are equal to $\omega_i^2/2$ are symmetric around $\tilde{\omega}^2$ and $\tilde{\epsilon}$ respectively, as shown in Fig.~\ref{fig10}(b). Due to this symmetry of $\omega_i^2$ around $\tilde{\omega}^2$ we immediately obtain that 
\begin{equation}
    \left| { \omega_{\frac{N+1}{2}}-\omega_{\frac{N+1}{2}-m}} \right| > \left| {\omega_{\frac{N+1}{2}}-\omega_{\frac{N+1}{2}+m}} \right|
\end{equation}

This asymmetry, shown in Fig.~\ref{fig10}(a), affects the  time evolution of the mass-spring chain,  since the important dynamical parameters are the eigenfrequencies $\omega_i$ (and not the eigenenergies) and also suggests that the excitations to the upper band's modes are higher than the excitations to the corresponding modes of the lower band.

\begin{figure}[h!]
\includegraphics[width=0.5\columnwidth]{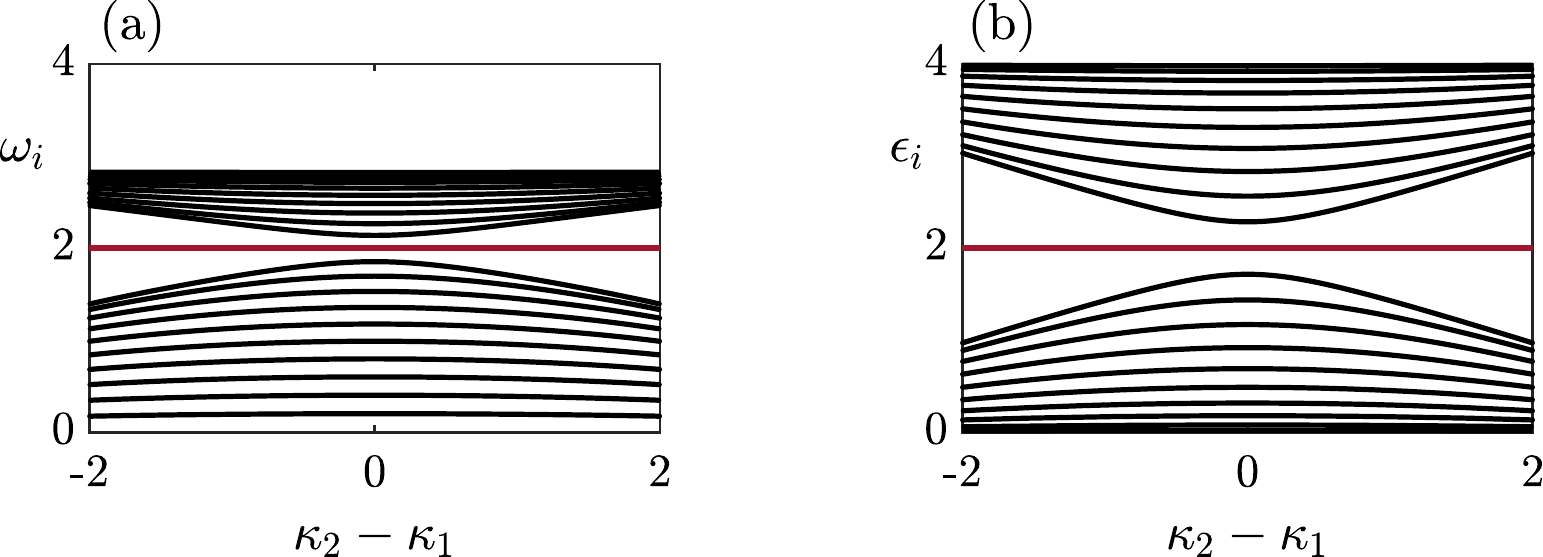}
\caption{(a) The frequencies $\omega_i$ are not symmetric around the edge mode’s frequency. (b) The eigenergies $\epsilon_i$ are symmetric around $\tilde{\epsilon}$.}
\label{fig10}
\end{figure}


\section{Oscillations of $F(\phi_0)$}
\label{appendixB}

The reason of defining the minimum of $F$ over all initial phases as fidelity is the following: The initial edge mode is oscillating, with period $\tilde{T}=2\pi/\tilde{\omega}$ and therefore we may choose any phase $\phi_0$ of this oscillation in order to determine the initial conditions, i.e., initial displacements and velocities. However, the quantity $F$, which measures the fraction of the system's final energy that is stored at the target edge mode, depends on the initial phase $\phi_0$. As an illustration, consider that the two couplings change in time through the 3-step up control scheme that is studied in the main text. In Fig.~\ref{fig11}(a) we show that for a fixed final time $T=15$, $F$ strongly oscillates as a function of the initial phase $\phi_0$. However, when the final time is $T=22$ [Fig.~\ref{fig11}(b)], $F$ is always above 0.99. Therefore, in order to ensure that the transfer has been achieved with a certain target fidelity ($99\%$ in this case), we consider the minimum of $F$ to be this fidelity. 

\begin{figure}[h!]
\includegraphics[width=0.5\columnwidth]{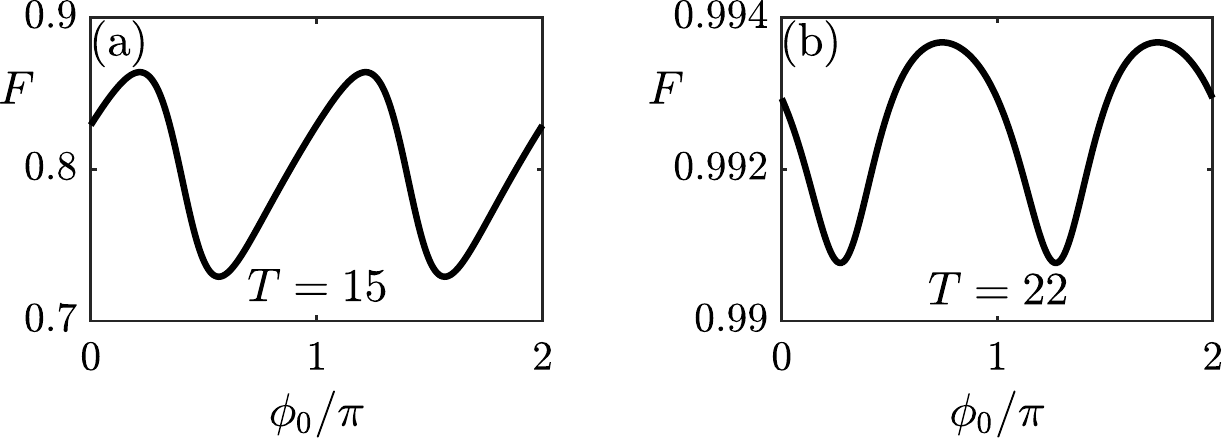}
\caption{$F$ as a function of the initial phase $\phi_0$ for the 3-step up protocol and for a final time $T=15$. (b) Same as (a) but for a final time $T=22$.}
\label{fig11}
\end{figure}

\end{widetext}


	

\begin{thebibliography}{99}
\bibitem{note1} Or are just independent from the off-diagonal.
\end{thebibliography}
\end{document}